\documentclass[superscriptaddress,amsmath,amssymb, apsq, pra]{revtex4-2}
\usepackage{bbold}
\usepackage{dsfont}
\usepackage{graphicx}
\usepackage{dcolumn}
\usepackage{caption}
\usepackage{subcaption}

\usepackage{multirow}
\usepackage{array}
\usepackage{bm}
\usepackage{amsmath}
\usepackage[colorlinks, allcolors={blue}]{hyperref}
\usepackage[T1]{fontenc}
\usepackage[dvipsnames]{xcolor}
\usepackage{comment}
\usepackage{braket}
\usepackage{relsize}
\usepackage{algpseudocode}
\newcounter{algorithm}

\begin{document}

\title{Quantum Error Mitigation with Diffusion-Like Models}

\newcommand{\affBIU}{Faculty of Engineering and the Institute of Nanotechnology and Advanced Materials, Bar-Ilan University, Ramat Gan 5290002, Israel}
\newcommand{\affNVIDIA}{NVIDIA, Hakidma 26, Ofer Industrial Park, Yokneam 2069203, Israel}
\newcommand{\affIQS}{Institute for Quantum Studies, Chapman University, Orange, California 92866, USA}
\newcommand{\affNVIDIAUS}{NVIDIA, Santa Clara, USA}

\author{Yuval Idan}
\affiliation{\affBIU}
\affiliation{\affNVIDIA}

\author{Ofek Nourian}
\affiliation{\affBIU}

\author{Elad Mentovich}
\affiliation{\affNVIDIA}

\author{Taylor L. Patti}
\affiliation{\affNVIDIAUS}

\author{Eliahu Cohen}
\affiliation{\affBIU}
\affiliation{\affIQS}

\begin{abstract}
Coupling between a quantum system and its environment is a fundamental decoherence mechanism, causing information encoded in the principal system to be progressively lost into environmental degrees of freedom. Such interactions can be modeled by a von Neumann-type system-environment coupling and, when resolved into discrete time steps, can be interpreted as a sequence of weak measurements that provides an effective model for noisy quantum dynamics.
Motivated by this picture, this work proposes an AI-assisted error-mitigation framework for quantum diffusion processes generated by sequential local weak measurements. Similar to classical diffusion models, the forward process progressively erases information from the input state. Here, this forward process is realized by sequential weak measurements that non-trivially drive the quantum state toward decoherence. 
Specifically, we model decoherence through sequential weak measurements in randomly chosen Pauli bases, which generate basis-dependent local dephasing and locally depolarizing dynamics on average. We subsequently train machine-learning models on exact synthetic density matrices to learn a channel- and distribution-specific denoising map and estimate the corresponding pre-noise quantum state.
The proposed approach is benchmarked using numerical experiments on single-qubit Bloch-vector states, separable multi-qubit registers, and entangled multi-qubit registers. A particular emphasis is given to the distribution-dependent  local-to-global reconstruction task, in which only local reduced density matrices are provided as input to reconstruct the full global density matrix. This setting is experimentally advantageous because it relies on locally accessible information rather than direct full-state tomography, making it more compatible with realistic noisy quantum devices and distributed quantum registers. Beyond state reconstruction, the proposed framework can also be viewed as part of a broader hybrid classical--quantum architecture, in which classical learning models are used to approximate non-unitary quantum dynamics and mitigate coherence loss across the system.

\end{abstract}

\maketitle

\section{Introduction}
Interactions between a quantum system, its environment, and external control mechanisms can be modeled using quantum channels and, more generally, process matrices \cite{coecke2017picturing, castro2018dynamics, tamir2013introduction, zurek2003decoherence}. These effective channels describe noise, imperfect control, and information exchange in quantum devices, including distributed quantum-computing architectures. Since incomplete characterization of such channels allows decoherence to accumulate in an uncontrolled manner, reliable quantum computation requires both improved channel understanding and active strategies for suppressing their effects.
Recent advances in quantum error correction, logical-qubit demonstrations, and below-threshold surface-code memories provide increasing evidence that the field may be approaching the first stages of fault-tolerant quantum computing \cite{paetznick2024logical,google2025belowthreshold,bluvstein2026fault,besedin2026lattice,putterman2025hardware}. Nevertheless, despite these advances, error mitigation remains an essential component of near-term and early fault-tolerant quantum computation, both because it can extend the feasible circuit depth of noisy devices and because it is well-suited for hybrid quantum-classical workflows \cite{aharonov2025importance,bharti2022noisy}.

Noise is deleterious to quantum processes as it interrupts their intended unitary evolution \cite{nielsen2010quantum}. The resulting noisy quantum channels can act asymmetrically on quantum systems, e.g., as bit-flips, phase-flips, and amplitude-damping. However, such random, asymmetric noise processes can, in aggregate, yield an effectively symmetric noise model known as the depolarizing channel \cite{nielsen2010quantum,leung2017complementary, king2003capacity}. Specifically, the depolarizing channel is \textit{isotropic}: it contracts all non-identity components uniformly and drives systems towards the maximally mixed state. Owing to these symmetric decoherence dynamics, the resulting process can be conceptualized as a forward process for quantum diffusion models \cite{ho2020denoising,nichol2021improved,sohl2015deep}. Such symmetric noise models also arise in quantum state reconstruction, including quantum tomography and classical shadows \cite{aaronson2018shadow,huang2020predicting,koh2022classical}. In these settings, positive operator-valued measure (POVM) measurement processes, together with device imperfections, induce decoherence. Moreover, randomized measurements and twirling often render the effective noise approximately depolarizing. To mitigate this type of noise, one typically characterizes the channel and applies its inverse, up to some level of approximation \cite{cai2023quantum}. However, the channel is not always known or identifiable with sufficient accuracy, making direct inversion impractical. More precisely, when the effective noise is only approximately depolarizing (i.e., when the true channel deviates from an ideal depolarizing map), characterizing it accurately enough for inversion becomes even more challenging. In such cases, channel-agnostic or robust mitigation methods are often preferable to exact inversion \cite{kandala2019error,strikis2021learning,czarnik2021error}.

In classical diffusion models, information is reconstructed by learning a reverse Markov chain that inverts a forward noising process, yielding a high-probability path back to the original data distribution \cite{Zhang2022DiffQAS,ho2020denoising}. By analogy, a quantum-diffusion-model approach can employ a sequence of learned denoising operations that approximate the inverse dynamics of the noisy process, transforming noisy mixed states into estimates of their corresponding pre-noise pure states~\cite{huang2020predicting,PhysRevLett.134.090801}.
This inverse is necessarily approximate and is typically realized via classical post-processing \cite{huang2020predicting,aaronson2018shadow}, variational circuits \cite{ravi2022vaqem}, or measurement-conditioned maps \cite{o2016efficient}, since the exact inverse of a noisy quantum channel is generally not physically implementable as a standalone quantum channel without additional classical or quantum resources \cite{resRevModPhys.91.025001}. In particular, such an inverse would generally reduce entropy and therefore need not be completely positive.

In this manuscript, we introduce diffusion-like models that effectively invert the effect of quantum noise channels by inferring the original pure state of a decohered quantum system. To generate such noisy trajectories, we randomly sample sequential weak measurements in the Pauli $x$, $y$, and $z$-bases \cite{aharonov1988result,hallaji2017weak,tamir2013introduction,troupe2017quantum}. Each realization is basis dependent, whereas uniform averaging over the three bases yields a locally depolarizing evolution. This construction retains microscopic variation between trajectories without assuming a globally depolarizing channel at every step. Due to such properties, weak measurements can serve as effective surrogates for iterative, distributed random noise processes. After generating noisy data via weak measurements, we train diffusion-like machine-learning models to act as an error-mitigation mechanism \cite{kandala2019error,temme2017error}. While in recent years machine-learning methods have become useful classical tools for supporting quantum-information tasks such as calibration, reconstruction, readout-error mitigation, and the analysis of noisy quantum data \cite{kim2020quantum,kim2022quantum,angelatos2021reservoir}, in the present work, we use neural networks in a novel and specific role: as effective inverse maps for non-unitary quantum channels. Likewise, we emphasize that we do not include any kind of quantum machine learning model, such as variational quantum circuits \cite{tilly2022variational,peruzzo2014variational,mcclean2016theory}. Our work proposes a classical machine learning model that infers quantum states from quantum measurement data.

Since our forward channel is decomposed into a sequence of weak-measurement steps, recurrent neural networks (RNNs) are natural candidates for learning inverse dynamics \cite{hochreiter1997long,chung2014empirical,sutskever2014sequence}. RNNs process sequential data by maintaining internal hidden states, and are therefore well suited to trajectory-based reconstruction tasks. The RNNs we employ in this work are long short-term memory (LSTM) models \cite{hochreiter1997long}, which we deploy on the linearly scaling separable datasets. In contrast, attention-based architectures, such as vision transformers \cite{wu2021cvt,touvron2021training,dosovitskiy2020image,wu2021cvt}, and enhanced U-Net architectures \cite{ronneberger2015u,isensee2021nnu,oktay2018attention}, with a custom attention mechanism,  are used for systems requiring non-separable (full density-matrix) representations, as they support parallel processing and can capture long-range correlations across the matrix. In the sequential setting, these correlations may also encode dependencies accumulated along the noisy trajectory.
For all datasets, the goal is to learn an effective denoising map for measurement-induced noisy trajectories, rather than to construct a physical inverse quantum channel. The forward process is generated by a noisy quantum channel composed of randomly sampled local weak-measurement maps, applied sequentially with a fixed interaction strength. For each representative trajectory, the weak-measurement sequence is applied until the state approaches a user-defined purity threshold. The resulting sequence defines an effective noisy channel, which is then applied to additional random initial quantum states. These states are not necessarily driven to the same final purity, since the effect of the channel depends on the input state. This procedure provides an approximate form of quantum-channel characterization \cite{chuang1997prescription,poyatos1997complete}, while the learning task is analogous to denoising in classical diffusion models \cite{tian2020deep,goyal2020image,xu2014dynamic,dabral2023mofusion}.

The importance of this approach is not limited to state reconstruction. Since the noisy evolution is generated by a sequence of local weak measurements, the learned object is closer to a measurement-induced quantum process than to a single fixed noise map. In this sense, the model learns an effective representation of contextual non-unitary dynamics, where the output depends not only on the initial density matrix, but also on the order, strength, and local structure of the weak measurements. This perspective is naturally related to process-matrix and process-tensor descriptions of open quantum systems, where quantum dynamics are characterized by multi-time correlations rather than by a single memoryless channel. Therefore, learning the inverse weak-measurement trajectory can be interpreted as a first step toward data-driven learning of contextual quantum processes \cite{coecke2017picturing}, with relevance to distributed quantum computing, quantum networks, and measurement-based quantum technologies\cite{kimble2008quantum,azuma2023quantum,raussendorf2001one,raussendorf2003measurement}.

\begin{figure}
    \centering
    \includegraphics[width=1.0\linewidth]{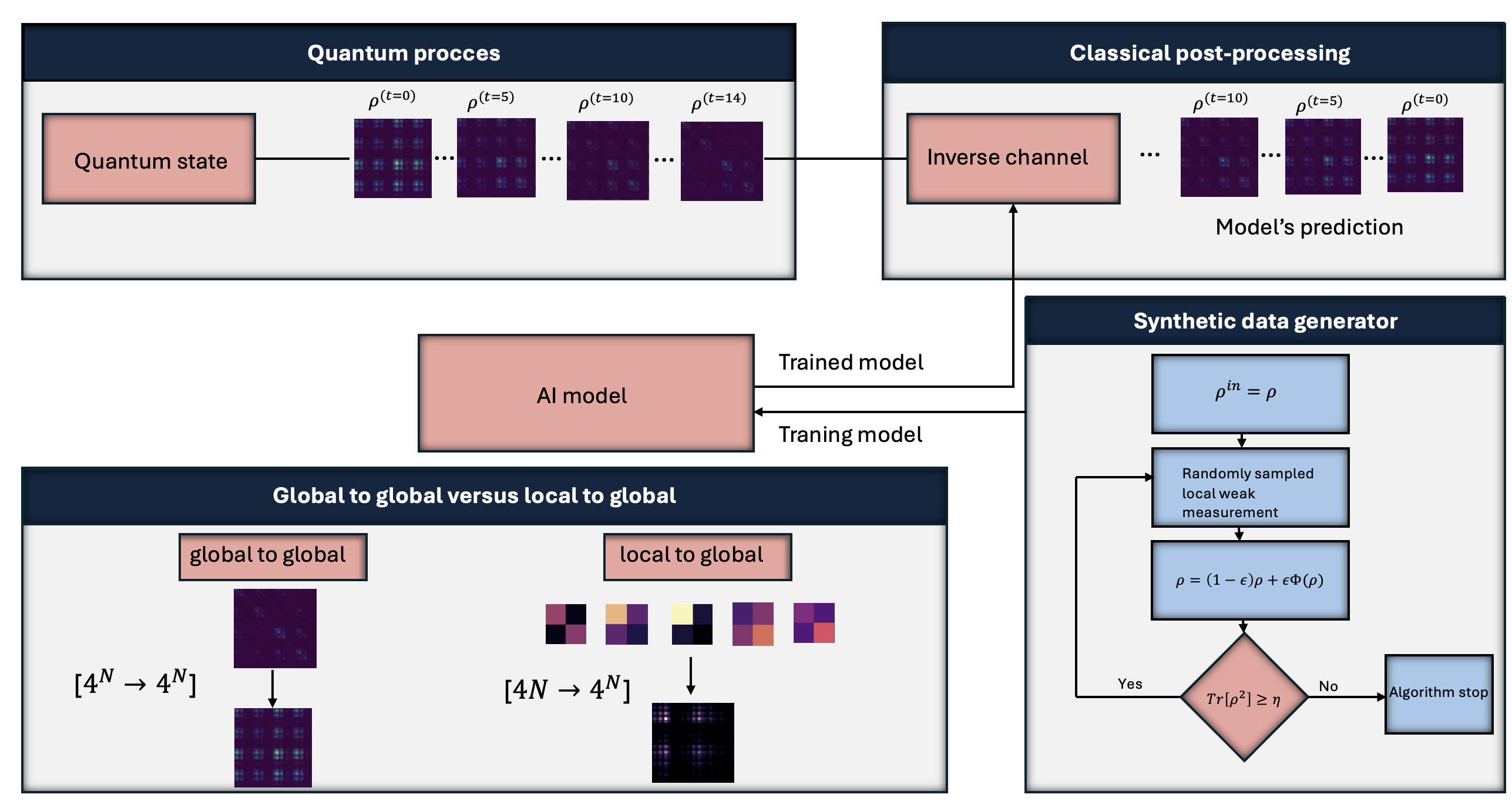}
    \caption{Schematic of the diffusion-like quantum-noise process and the AI-assisted inverse-channel reconstruction. An initial quantum state $\rho_{0}$ is evolved through a sequence of weak-measurement-induced noisy channels, producing a trajectory of mixed states $\{\rho_{t}\}_{t=1}^{T}$. An AI model is trained to approximate the inverse channel and map each noisy intermediate state $\rho_{t}$ back toward the original state $\rho_{0}$. The upper panels show the forward noisy evolution and the reverse trajectory predicted by the trained model. The lower-left panel compares global-to-global reconstruction, in which the complete noisy $N$-qubit density matrix is provided, with local-to-global reconstruction, in which only the single-qubit reduced density matrices are used to reconstruct the initial global entangled state. The lower-right panel shows the synthetic-data-generation pipeline, where randomly sampled local weak measurements are iteratively applied until a predefined purity threshold is reached.}
    \label{fig:paper's_pipline}
\end{figure}

The remainder of this manuscript is organized as follows. In Sec.~\ref{sec:weak-mesaurement} and Sec.~\ref{Sec:diffusionmodel}, we introduce the formalisms of weak-measurement and diffusion models, respectively. In Sec.~\ref{subsec:qdm} and Sec.~\ref{sec:forward_process}, we describe our quantum diffusion-like models and the forward process we use to generate synthetic noisy trajectory training data from sequential weak measurements (the overall workflow of these subsections is summarized in Fig.~\ref{fig:paper's_pipline}). Sec.~\ref{Sec:data_sets} presents the datasets studied in this work: single-qubit Bloch-vector diffusion, separable multi-qubit registers, entangled registers with local noise, and local-to-global reconstruction from reduced density matrices. Across these settings, LSTM, U-Net, and Vision Transformer architectures are explored. Finally, in Sec.~\ref{sec:dis}, we discuss the results, limitations, and future directions.

\section{Theoretical Background}
\subsection{Sequential weak measurements}\label{sec:weak-mesaurement}

When a quantum state $\ket{\psi}=\sum_i \alpha_i\ket{a_i}$, where $\sum_i|\alpha_i|^2=1$, is strongly measured, it is projected onto the eigenbasis $\{\ket{a_i}\}$ of the target observable $A$. Consequently, the measurement outcome contains information only about that observable, while information about non-commuting observables is erased. Conversely, we can consider a trade-off between the accuracy of the information obtained and the disturbance induced on the wave function. A measurement that collapses only part of the wave function is called a weak measurement, since the interaction applied to the quantum state is correspondingly weak. This procedure can be generalized to a continuous spectrum of increasing interaction strengths, in the limit of which, we recover traditional strong quantum measurement. In this section, the mathematical background and formalism of weak measurement are presented.

Without loss of generality, let us measure the quantum state $\ket{\psi}$ in the eigenbasis of the observable $A$. The state $\ket{\psi}$ is weakly coupled to an auxiliary quantum system known as the pointer, which is a continuous-variable quantum system, $\ket{\phi} = \int \psi(x)\ket{x}\,dx,$
where $\int |\psi(x)|^2 dx = 1$. A von Neumann measurement interaction ~\cite{tamir2013introduction} between the state and the pointer results in the state
\begin{equation}\label{Basic_weak_meas}
e^{-i\gamma A\otimes P_d}\ket{\psi}\ket{\phi(x)}
=
\sum_i \alpha_i \ket{a_i}\ket{\phi(x-\gamma a_i)}.
\end{equation}
Here, $\gamma$ represents the coupling parameter, which is assumed to be small ($\gamma \ll 1$), and $P_d$ denotes the momentum operator of the pointer. The pointer shifts according to each eigenvalue $a_i$ proportional to the interaction strength $\gamma$. Tracing out the pointer leads to the monitoring transformation~\cite{dieguez2018information}:
\begin{equation}\label{eq:monitoring}
    \mathcal{M}(\rho)=(1-\epsilon)\rho +\epsilon\Phi_{A}(\rho),
\end{equation}
where $\rho=\ket{\psi}\bra{\psi}$, $\epsilon$ is the phenomenological measurement strength, and, for the non-selective dephasing model used below, $\Phi_A(\rho)=\sum_a P_a\rho P_a$ with $P_a$ the eigenspace projectors of $A$. Equation~\ref{n_monitors} assumes this projective-dephasing channel, so that $\Phi_A^2=\Phi_A$.

The above superoperator describes the transformation of the quantum state under measurement. In particular, it decomposes the measured state into its initial density matrix and measurement-projected components, the latter of which is in the eigenbasis of the measured observable. The partition between the pre-measurement component and the projected component is determined by the measurement strength, which is parameterized by $\epsilon$. If the weak measurement is performed in the eigenbasis of an observable that does not commute with the density matrix, the purity of the state decreases. In other words, the state accumulates noise, which results in information loss.

Repeating weak measurements of the same observable approaches the corresponding projective dephasing channel only in the many-repetition limit, while at finite sequence length a residual coherent component remains. When the measurement record is retained, such a sequence can provide partial information about several observables. In the simulations below, however, the outcomes are discarded and only the induced non-selective disturbance is modeled. The weaker the measurement, the lower the signal-to-noise ratio of any retained record becomes. Likewise, sequential weak measurements may disturb the quantum state more than a single projective measurement due to the presence of multiple non-commuting observables, each of which may erase information about the others.

Performing sequential weak measurements of the same observable with the same measurement-strength parameter $\epsilon$ leads to the iterative transformation \cite{dieguez2018information}
\begin{equation} \label{n_monitors}
    \mathcal{M}^n(\rho)=(1-\epsilon)^n\rho + \bigl(1-(1-\epsilon)^n\bigr)\Phi_{A}(\rho).
\end{equation}
From the above equation, we see that the state is gradually transformed into its projected components in the observable's eigenbasis. Performing sequential weak measurements of many different observables $\mathcal{O}_i\in \{\mathcal{O}_1, \mathcal{O}_2, \dots , \mathcal{O}_n\}$ yields
\begin{equation}    \label{Eq:alot_of_weak}
    \rho^{\mathrm{out}} = \mathcal{M}_{\epsilon}^{\mathcal{O}_n} \circ \cdots \circ \mathcal{M}_{\epsilon}^{\mathcal{O}_1}(\rho),
\end{equation}
where $\mathcal{M}_{\epsilon}^{\mathcal{O}_i}$ is the monitoring quantum channel with interaction strength $\epsilon$ in the eigenbasis of $\mathcal{O}_i$. Note that
\begin{equation}
\mathcal{M}_{\epsilon}^{\mathcal{O}_i}\mathcal{M}_{\delta}^{\mathcal{O}_j}(\rho)
= (1-\epsilon)(1-\delta)\rho +(1-\epsilon)\delta \Phi_{\mathcal{O}_j}(\rho) +(1-\delta)\epsilon \Phi_{\mathcal{O}_i}(\rho) + \epsilon \delta \,\Phi_{\mathcal{O}_i}\!\bigl(\Phi_{\mathcal{O}_j}(\rho)\bigr).
\label{eq.different_observables}
\end{equation}
Equation~\ref{eq.different_observables} shows that the weights of the ordered components depend on the measurement strengths and on the order of the generally non-commuting dephasing maps. With such a sequence of weak measurements, the state becomes increasingly mixed and may approach the maximally mixed state, leading to information loss. However, due to the spectral decomposition and the causal order of the measurements, this outcome does not generally occur in a trivial manner. In addition, because the output density matrix is distributed across many observables, finding an inverse channel in a high-dimensional Hilbert space may be a complicated task.

\subsection{Diffusion Models}\label{Sec:diffusionmodel}

Classical diffusion models~\cite{ho2020denoising} are Markovian processes in which information is gradually erased through the accumulation of noise, often modeled as Gaussian noise. Such models consist of a forward noising process and a learned reverse denoising process~\cite{ho2020denoising,sohl2015deep}. The forward trajectory acts on the data distribution as
\begin{equation}
    q(x_{1:T}|x_0) = \prod_{t=1}^{T}q(x_t|x_{t-1}).
\end{equation}
The reverse process is parameterized by learnable functions and written as

    \begin{equation}
 p_{\theta}(x_{0:T}) = p(x_{T})\prod_{t=1}^{T}p_{\theta}(x_{t-1}|x_{t}), 
 \end{equation}

where
\[
p_{\theta}(x_{t-1}|x_{t}) = \mathcal{N}(x_{t-1};\mu_{\theta}(x_t,t),\Sigma_{\theta}(x_t,t)),
\qquad
q(x_t|x_{t-1}) = \mathcal{N}(x_t;\sqrt{1-\beta_t}\,x_{t-1},\beta_t I).
\]
Here, $p_{\theta}(x_{0:T})$ is the joint distribution associated with the reverse process, which is a Markov chain with learned Gaussian transitions, and $q(x_{1:T}|x_0)$ is the diffusion process (forward process). In addition, $p(x_T)=\mathcal{N}(x_T;0,I)$ is the Gaussian prior, and $\beta_1,\ldots,\beta_T$ is the variance schedule of the diffusion process.
\section{Methods}
\subsection{Quantum diffusion model}
\label{subsec:qdm}
In this work, we use the diffusion-model structure as inspiration for a quantum analogue that characterizes dissipative quantum processes. The forward process is formulated as a $T$-step quantum Markov process generated by sequential weak measurements. In the implementation studied here, each local step samples one of the three Pauli measurement bases, and applies the corresponding non-selective partial-dephasing channel. Acting sequentially on the quantum state, these measurements generate a diffusion-like evolution that gradually erases information from the initial state and drives the system toward a highly mixed state. In this framework, the interaction strength between the pointer and the system state plays a role analogous to the noise parameter $\beta$ in classical diffusion models. Random sequential weak measurements drawn from a measurement ensemble induce a non-unitary noisy evolution. When the measurements are sampled symmetrically over the relevant observables, the averaged dynamics can exhibit depolarizing-like behavior. Therefore, the process may be approximated by a depolarizing quantum channel~\cite{nielsen2010quantum}, defined as
\begin{equation}\label{Eq._depolorazingqchannel}
    D_{p}(\rho) = p\rho + (1-p)\frac{\text{Tr}(\rho)}{2^n}\mathds{I},
\end{equation}
where $\mathds{I}$ is the identity operator and $2^n$ is the Hilbert-space dimension of the density matrix. This process resembles the forward process of classical diffusion models, where here $1-p$ plays the role of the Gaussian noise parameter in the classical setting.

The formal inverse of the depolarizing quantum channel is defined as~\cite{nielsen2010quantum}
\begin{equation}
    D^{-1}_{p}(\rho)=\frac{1}{p}\rho +\left(1-\frac{1}{p}\right)\frac{\text{Tr}(\rho)}{2^n}\mathds{I}.
\end{equation}
For $0<p<1$, the above expression is a formal linear inverse of the depolarizing map, but it is not strictly positive, and hence not a completely-positive trace-preserving (CPTP) map on the full state space. A physical reversal would require additional information, such as access to the environment in a Stinespring dilation, side information, or a restricted correctable subspace. Note that in the present work the noise channels are local, meaning that each weak-measurement map acts independently on a single qubit. Therefore, a product of local depolarizing-like channels is not generally equivalent to a single global depolarizing channel, i.e., $\left(\bigotimes_{i=1}^{n}D^{(i)}_{p_i}\right)(\rho) \neq D_p(\rho)$.
The learned mitigation map should therefore be understood as an effective inverse map on the sampled noisy trajectories, rather than as a physical inverse quantum channel. Formally, the ideal reconstruction relation can be written as
\begin{equation}
\rho \approx \mathcal{M}^{-1}_{\mathrm{eff}}(\mathcal{M}(\rho)).
\end{equation}

Here, we introduce a quantum analogue of the classical diffusion model. In traditional diffusion models, the forward process, $q(x_{1}\vert x_0)$, describes the accumulated noise generated by a Markov chain, while the inverse process, $p_{\theta}(x_{0})$, is a parametrized denoising process, which reconstructs the original quantum state. In the quantum diffusion-like models that we propose, this corresponds to a sequence of noise-inducing quantum channels followed by a learned inverse map that transforms noisy mixed states into estimates of their corresponding pre-noise pure states. Both the forward and inverse channels can be decomposed into a sequence of discrete time steps, each associated with a weak measurement. The action of each such step on the input quantum state is given by the monitoring quantum channel defined in Eq.~\ref{eq:monitoring}. Consequently, the entire forward process, consisting of sequential weak measurements, may be written as
\begin{equation}\label{diffusion_procces}
    \mathcal{M}^{(1\ldots T)}\left(\rho\right)
    =
    \mathcal{M}^{(T)} \circ \mathcal{M}^{(T-1)} \circ \cdots \circ \mathcal{M}^{(1)}(\rho).
\end{equation}

As discussed previously, individual trajectories are basis dependent, whereas the ensemble-averaged dynamics are locally depolarizing. That is, at each time step, the dynamics consist of local quantum channels acting independently on each Hilbert space,
\begin{equation}
\mathcal{M}^{(t)} = \bigotimes_{i=1}^{n} \mathcal{M}^{(t)}_i,
\label{eq:local_channels}
\end{equation}
with action
\begin{equation}
    \rho^{(t)} = \mathcal{M}^{(t)}(\rho^{(t-1)}) = \left(\bigotimes_{i=1}^n \mathcal{M}_{i}^{(t)} \right) (\rho^{(t-1)}).
\end{equation}
Here, $\mathcal{M}^{(t)}_i$ denotes the monitoring quantum channel acting on the $i$th subsystem at time step $t$. For an $n$-qubit register, the global Hilbert-space dimension is $2^n$, while each local weak-measurement map acts on a single-qubit subsystem. In this work, such a decomposition is assumed since the weak measurements act independently on each qubit. For $\mathcal{M}^{\epsilon}_c=(1-\epsilon)\mathds{I}+\epsilon\Phi_c$ and uniform sampling over $c\in\{x,y,z\}$, the ensemble-averaged single-qubit step is
\begin{equation}\label{eq:pauli_average}
\overline{\mathcal{M}}^{\epsilon}=\frac{1}{3}\sum_{c=x,y,z}\mathcal{M}^{\epsilon}_c=D_{1-2\epsilon/3}.
\end{equation}
Accordingly, independent local sampling on an $n$-qubit register yields $\bigotimes_{i=1}^{n}D^{(i)}_{1-2\epsilon/3}$ on average, rather than a single global depolarizing channel. Subsequently, the forward process is defined in Eq.~\ref{diffusion_procces}, where $T$ denotes the total number of discrete time steps. A formal stepwise inverse, when it exists algebraically, is given by
\begin{equation}\label{eq:inverse}
\mathcal{M}_{\mathrm{inv}}^{(T)}
=
(\mathcal{M}^{(1)})^{-1} \circ (\mathcal{M}^{(2)})^{-1} \circ \cdots \circ (\mathcal{M}^{(T)})^{-1}.
\end{equation}
Table~\ref{tab:diffusion} summarizes this relationship between the classical diffusion model in~\cite{ho2020denoising} and the proposed quantum version.

\begin{table}[h!]
    \centering
    \caption{Diffusion model: the analogy between the classical model in \cite{ho2020denoising} and the suggested quantum model.}
    \label{tab:diffusion}
    \begin{tabular}{|l||l|l|}
        \hline
        \multicolumn{3}{|c|}{Diffusion model} \\
        \hline
        Attribute & classical & quantum \\
        \hline
        Forward  & $q(x_{t} \mid x_{t-1})$ & $\rho^{(t)} = \mathcal{M}(\rho^{(t-1)})$ \\
        Reverse &  $p_{\theta}(x_{t-1} \mid x_{t})$ & $\hat\rho^{(t-1)}=\mathcal{R}^{(t)}_{\theta}(\hat\rho^{(t)})$ \\
        Noise    & $\beta$ & $\epsilon$ \\
        Map      & Gaussian noise & Pauli-basis dephasing, depolarizing on average \\
        \hline
    \end{tabular}
\end{table}

\subsection{Implementation of the forward process}\label{sec:forward_process}

This section explains how the synthetic data for the non-unitary evolution is generated by the forward process introduced in Eq.~\ref{diffusion_procces}. In the implemented simulations, each local step samples a Pauli basis $c\in\{x,y,z\}$ and applies the non-selective weak-dephasing channel $\Phi_c(\rho)=\sum_{s=\pm}P_{s,c}\rho P_{s,c}$. The averaged map is therefore the locally depolarizing channel in Eq.~\ref{eq:pauli_average}, while individual trajectories remain basis dependent. This randomization is closely related to weak Pauli twirling \cite{xu2025efficientmeasurementerrormitigation}.

The algorithm proceeds to apply random weak measurements independently on each local Hilbert space until the density matrix reaches the prescribed purity criterion. The resulting quantum channel is then recorded and applied to many additional random states sampled according to the Haar measure on $SU(2)$. The overall procedure is illustrated in Fig.~\ref{fig:paper's_pipline} (under synthetic data generator), while the explicit algorithm used to generate the random channel is given in Algorithm~\ref{alg:forward}. The resulting dataset is then partitioned into training and validation sets, which are used to evaluate the performance of the trained model. 

\begin{center}
\refstepcounter{algorithm}\label{alg:forward}
\begin{minipage}{0.95\linewidth}
\noindent\textbf{Algorithm~\thealgorithm. Forward process}
\begin{algorithmic}
\State $u \gets \mathrm{rand}[0,1]$, $\theta \gets \arccos(1-2u)$, $\phi \gets \mathrm{rand}[0,2\pi]$ \Comment{Haar-uniform single-qubit state}
\State $\ket{\psi} \gets \cos(\theta/2)\ket{0} + \sin(\theta/2)e^{i\phi}\ket{1}$
\State $\rho \gets \ket{\psi}\bra{\psi}$
\While{$\mathrm{Tr}(\rho^2) \geq P_{\mathrm{stop}}$}
    \State $c \gets \mathrm{rand}(x,y,z)$ \Comment{Randomly choose a measurement basis}
    \State $\rho \gets (1-\epsilon)\rho + \epsilon \Phi_c(\rho)$, $\quad \Phi_c(\rho)=\sum_{s=\pm}P_{s,c}\rho P_{s,c}$
    \State $\rho \gets \rho / \mathrm{Tr}(\rho)$ \Comment{Numerical safeguard for trace drift}
\EndWhile
\end{algorithmic}
\end{minipage}
\end{center}

For the single-qubit local-channel generator in Algorithm~\ref{alg:forward}, $P_{\mathrm{stop}}=\eta$ ($Tr(\rho^2)= \eta$); for a global $n$-qubit density matrix, the analogous maximally mixed purity threshold is $2^{-n}$.

A characteristic feature of sequential weak measurements is the gradual emergence of mixed components associated with different measurement directions. As a result, the evolution does not approach the maximally mixed state in a trivial manner. Moreover, as the state becomes more mixed, driving it even closer to the maximally mixed state becomes increasingly difficult. In particular, for small disturbance strength $\epsilon$, the required sequence length may become very large. Therefore, in the examples presented here, the evolution is terminated within a feasible purity range. This limitation becomes especially significant for global states of many qubits, where reaching a state arbitrarily close to the global maximally mixed state is generally impractical.
All experiments were performed on an NVIDIA RTX 5090 GPU (32 GB, Blackwell architecture).

\section{Results}\label{Sec:data_sets}

This section describes the synthetic datasets generated via the forward process of Sec.~\ref{sec:forward_process}, together with the corresponding supervised learning tasks. All datasets are constructed from noisy trajectories produced by sequential local weak measurements, but they differ in the representation of the quantum state and in the amount of information available to the model.

The reconstruction tasks are ordered by increasing representational complexity. The first is single-qubit reconstruction using the Bloch vector. The second extends this setting to separable multi-qubit registers, for which local density matrices can be processed independently and the feature count scales linearly with the number of qubits. The third treats entangled registers, whose full density matrices encode nonlocal correlations and scale exponentially with system size. Finally, the local-to-global task uses only local reduced density matrices to predict the full global state. This setting is experimentally motivated by the greater accessibility of local measurements, but it is distribution dependent: for arbitrary quantum states, one-qubit marginals do not uniquely determine the global density matrix. The network therefore learns a reconstruction rule induced by the specific state-preparation and weak-measurement process, not a universal map from local marginals to global states. For the LSTM tasks, each local Hermitian $2\times2$ density matrix is encoded by four real parameters. For the image-based ViT and U-Net tasks, a global $d\times d$ complex density matrix, with $d=2^n$, is encoded as two real-valued channels containing its real and imaginary parts. As noted above, the models used in this work are LSTM, ViT, and an enhanced U-Net. Further model details are provided in Appendix~\ref{app:models}. Table~\ref{tab:data_sets} summarizes the implemented dataset types and the models used to handle them.

\begin{table}[h!]\label{table:data_sets}
    \centering
    \caption{Summary of the datasets and corresponding learning tasks.}
    \label{tab:data_sets}
    \begin{tabular}{ |l||l|l|l| }
        \hline
        Type & Model & Input & Output \\
        \hline
        Bloch vector & LSTM & $[B,S,3]$ & $[B,S,3]$ \\
        \hline
        Separable register & LSTM & $[B,S,4n]$ & $[B,S,4n]$ \\
        \hline
        Entangled register & ViT/U-Net & $[B,S,2,d,d]$ & $[B,S,2,d,d]$ \\
        \hline
        Local-to-global & ViT/U-Net & $[B,S,n,2,2,2]$ & $[B,S,2,d,d]$ \\
        \hline
    \end{tabular}
\end{table}

\begin{figure}[t]
    \centering
    \includegraphics[scale=0.4]{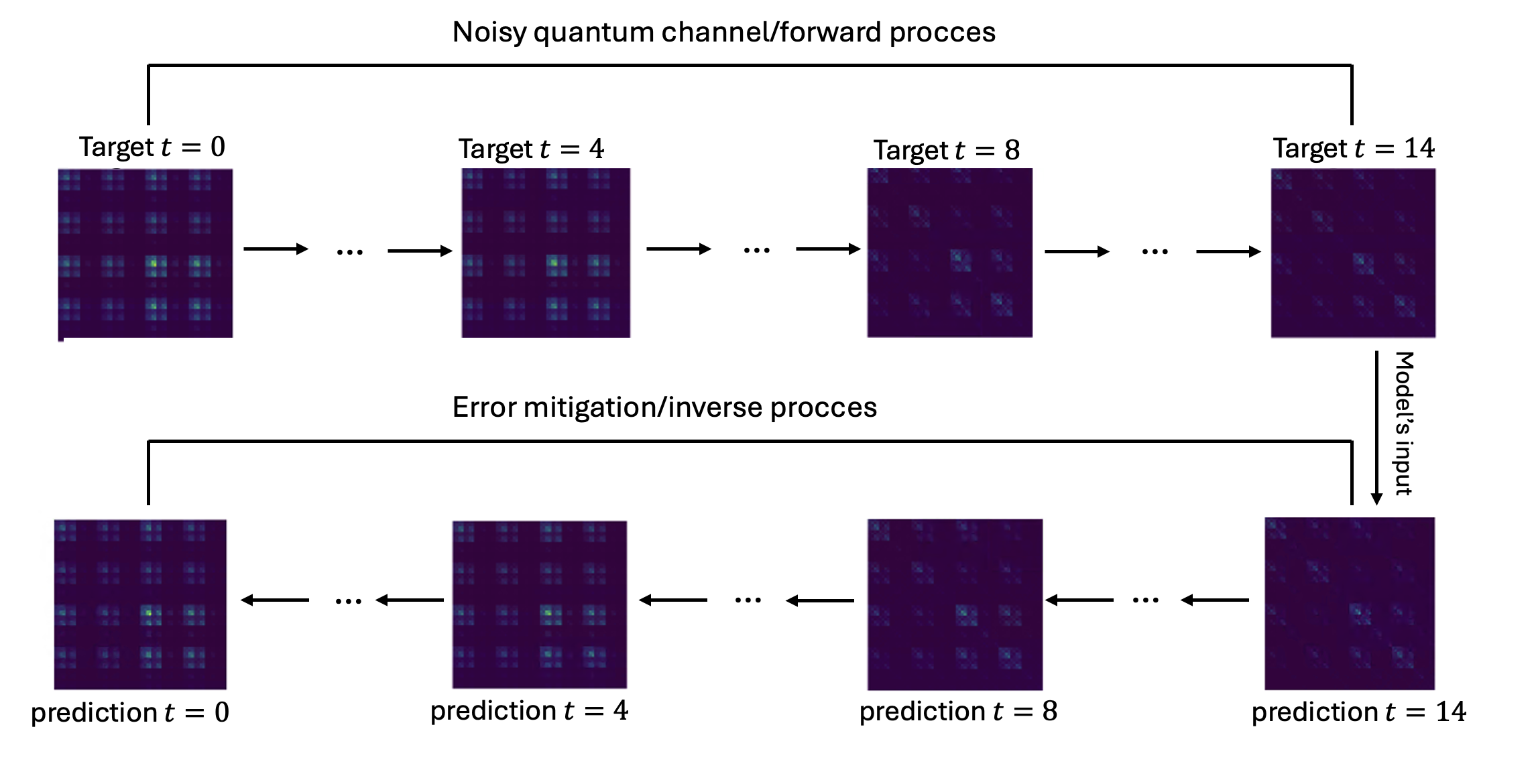}
    \caption{Example of the separable six-qubit denoising task. An input state evolves under a noisy quantum channel and is gradually degraded. The trained model receives the noisiest state, corresponding to the output of the noisy channel, and learns an effective inverse map that reconstructs the erased information. The reconstructed state achieves $\mathcal{F}_{GM}>0.99$ for a sequence length of $15$ steps.  }
    \label{fig:6 qubits}
\end{figure}

The reconstruction is guided primarily by the fidelity between the predicted
state $\hat{\rho}$ and the target ground-truth state $\rho$. To additionally suppress local
deviations between their individual matrix elements, a small L1 ($\mathcal{L}_{\mathrm{L1}}=\sum_{i=1}^{N} | y_i - x_i| $, where $x_i$ and $y_i$ denote the predicted and ground-truth features, respectively)  contribution  is
included, giving the loss function
\begin{equation}
    \mathcal{L}
    =
    \lambda_F\left[1-\mathcal{F}_{GM}(\hat{\rho},\rho)\right]
    +
    \lambda_1\mathcal{L}_{\mathrm{L1}},
\end{equation}
with $\lambda_F=10$ and $\lambda_1=10^{-2}$. Here,
\begin{equation}
    \mathcal{F}_{GM}(\hat{\rho},\rho)
    =
    \frac{\mathrm{Tr}(\hat{\rho}\rho)}
    {\sqrt{\mathrm{Tr}(\hat{\rho}^{2})
    \mathrm{Tr}(\rho^{2})}}
\end{equation}
is the geometric-mean fidelity \cite{liang2019quantum}, which quantifies the
overlap between the reconstructed and target density matrices. The much
smaller weight assigned to the L1 term reflects its secondary role: it
regularizes element-wise deviations without competing with the fidelity as
the principal reconstruction criterion.
The validation fidelity, evaluated using the exponential-moving-average
weights, determines model selection, learning-rate reduction, and early
stopping. Accordingly, fidelity is also used as the principal measure of the
reconstruction quality reported on the test set.

\begin{figure}
\centering
\includegraphics[width=1\linewidth]{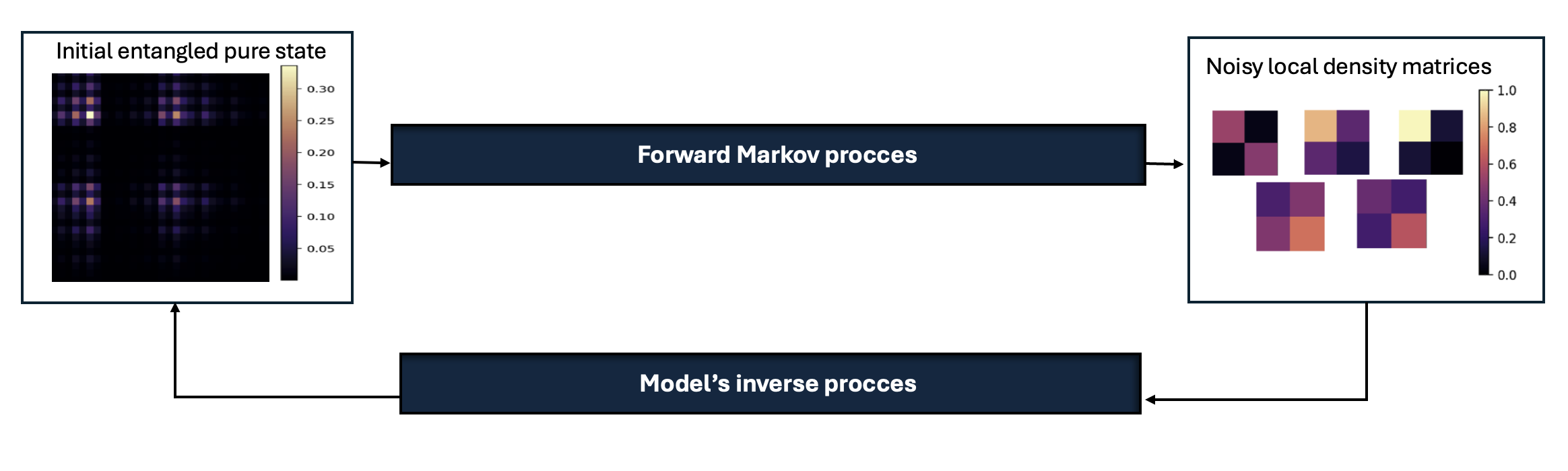}
\caption{Example of the local-to-global inverse reconstruction task for a five-qubit quantum register, which constitutes the most challenging dataset considered in this work. The reconstruction is performed using the Vision Transformer model. The model receives as input the five single-qubit reduced density matrices obtained from the noisy global state and predicts the corresponding initial global density matrix.}
\label{fig:local_to_global}
\end{figure}

\subsection{Non-interacting qubits}
\subsubsection{Single-qubit Bloch-vector dataset}\label{Sec:Bloch_vectors}

We begin with the simplest reconstruction task: the inverse evolution of a single qubit represented by its Bloch vector,
\begin{equation}
\rho = \frac{1}{2}(I+\vec{r}\cdot\vec{\sigma}),
\end{equation}
where $\vec{r_i}=(r_i^x,r^y_i,r^z_i)$ contains three real features per time step. This representation is useful for building intuition, since the noisy evolution can be visualized directly as a trajectory inside the Bloch sphere.

For each training example, a random initial pure state is generated and evolved under the forward process until a prescribed purity threshold is reached. The resulting trajectory is recorded as
\begin{equation}
\vec{r}_i =
\left(
\vec{r}^{(0)}_i,
\vec{r}^{(1)}_i,
\ldots,
\vec{r}^{(T)}_i
\right),
\end{equation}
where $i$ labels the training example and the superscript denotes the time step. The same sampled noisy channel is then applied to additional random initial states to generate the synthetic dataset. The input to the model therefore has shape
\begin{equation}
[B,S,3],
\end{equation}
where $B$ is the batch size and $S$ is the sequence length. In this representation, increasing noise corresponds to a reduction of the Bloch-vector norm, $|\vec{r}|_2 \rightarrow 0$.

The noisy single-qubit evolution can be written as an affine map,
\begin{equation}
\vec{r}^{(t+1)} = M\vec{r}^{(t)}+\vec{b},
\end{equation}
and, formally, the corresponding inverse map is
\begin{equation}
\vec{r}^{(t)} = M^{-1}\left(\vec{r}^{(t+1)}-\vec{b}\right).
\end{equation}
For weak disturbances, the forward step may be approximated as
\begin{equation}
\vec{r}^{(t+1)}
=
(1-\epsilon)\vec{r}^{(t)}
+
\epsilon\left(M\vec{r}^{(t)}+\vec{b}\right).
\end{equation}
Because the underlying evolution is non-unitary, the inverse map is generally not a physical quantum channel on the original Hilbert space. That is, our model learns an effective inverse map from data rather than constructing an explicit physical inverse process. The evolution of a single Bloch vector was readily reconstructed using an LSTM, demonstrating that the model could efficiently learn the comparatively simple single-qubit dynamics and providing a baseline for the more challenging and physically relevant datasets considered subsequently.

\begin{figure}[htbp]
\centering

\begin{subfigure}[t]{0.57\linewidth}
    \centering
    \includegraphics[width=\linewidth]{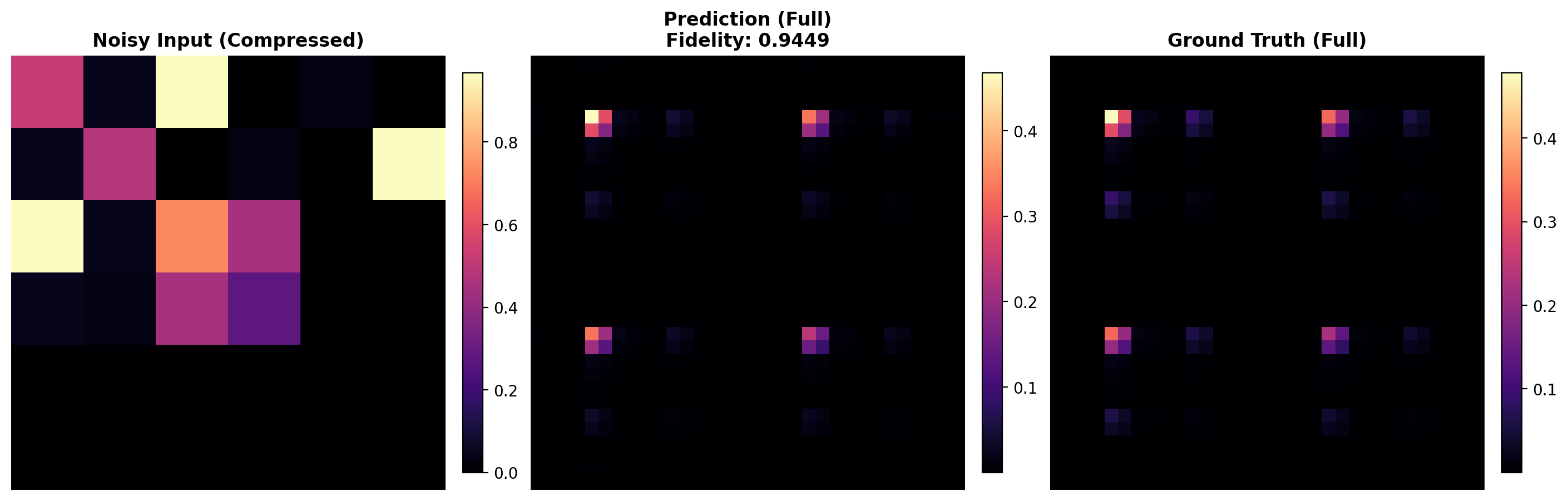}
    \caption{}
    \label{fig:sample_loc}
\end{subfigure}
\hfill
\begin{subfigure}[t]{0.39\linewidth}
    \centering
    \includegraphics[width=\linewidth]{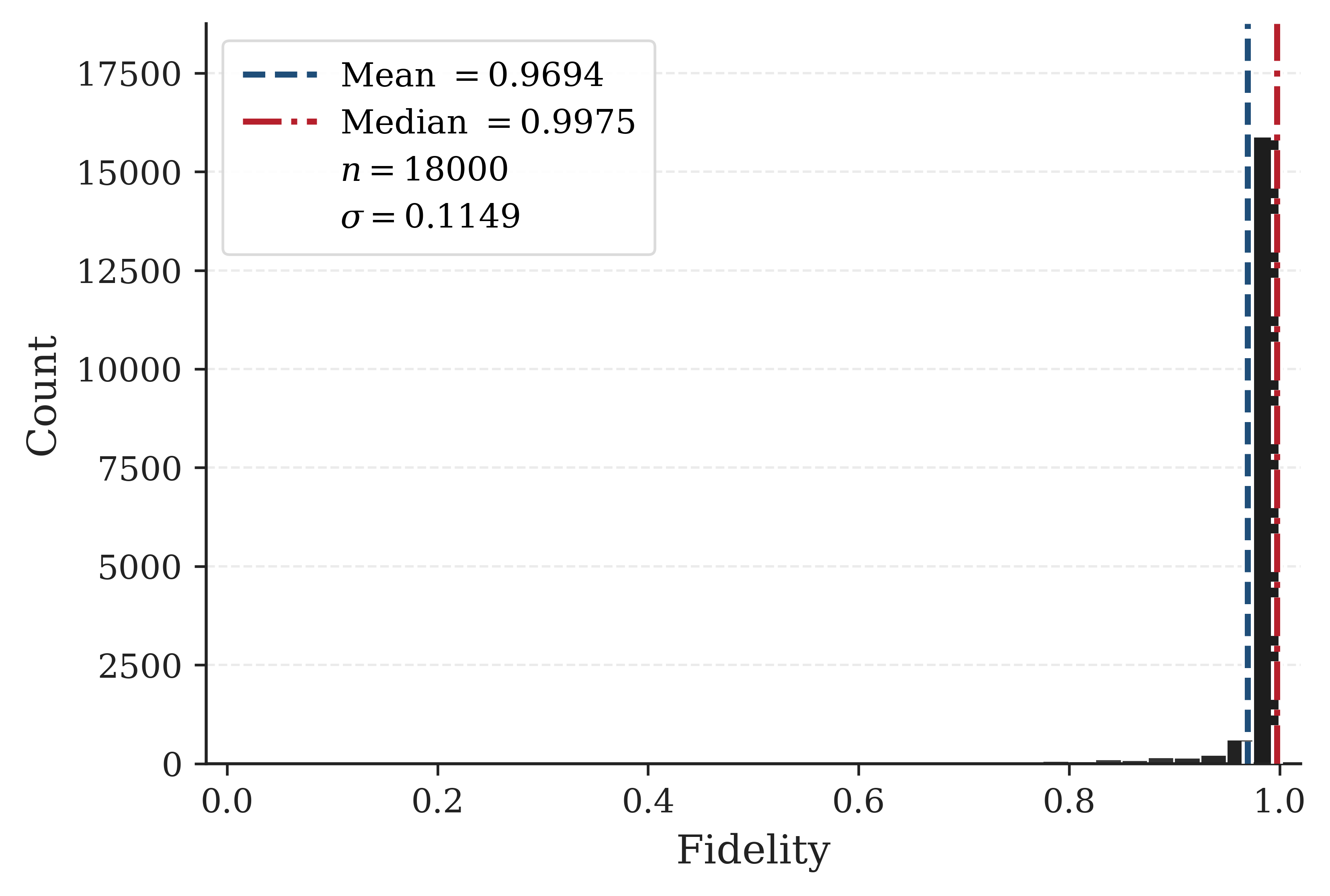}
    \caption{ }
    \label{fig:test_fidelity_hist}
\end{subfigure}

\captionsetup{
    width=\textwidth,
    font=normalsize,
    justification=justified,
    singlelinecheck=false
}

\caption{
Configuration and local-to-global reconstruction results. (a)~Example of a noisy local quantum-register input, together with the model prediction and the corresponding ground truth. The reconstructed state achieves $\mathcal{F}_{GM}=0.9449$. (b)~Reconstruction-$\mathcal{F}_{GM}$ distribution on the held-out test set ($n=18{,}000$ states). The mean and median fidelities are $\overline{\mathcal{F}}_{GM}=0.9694$ and $\widetilde{\mathcal{F}}_{GM}=0.9975$, respectively, with standard deviation $\sigma=0.1149$. Moreover, $91.6\%$ of the states achieve $\mathcal{F}_{GM}\geq0.95$, while $80.0\%$ achieve $\mathcal{F}_{GM}\geq0.99$. The model is trained on $n=5$-qubit states generated with $m=4$, $\epsilon=0.1$, and a target purity of $0.55$. Each of the $15{,}000$ generated states is expanded into a noisy trajectory of sequence length $13$. The data are split $80/10/10$ into $12{,}000$ training, $1{,}500$ validation, and $1{,}500$ test base states, corresponding to $18{,}000$ noisy--clean test pairs. Training uses AdamW with a learning rate of $5\times10^{-5}$, weight decay of $10^{-3}$, a batch size of $300$, and $500$ epochs, together with mixed-precision training, gradient-norm clipping at $1.0$, EMA weight averaging with $\tau=0.999$, and projection of the outputs onto valid density matrices. The best model, obtained at epoch $415$, achieves a validation $\mathcal{F}_{GM}=0.9653$, a mean test $\mathcal{F}_{GM}=0.9694$, and a median test $\mathcal{F}_{GM}=0.9975$.}\label{fig:test_fidelity}
\end{figure}

\subsubsection{Separable multi-qubit dataset}\label{sec:qregister}
We next consider the extension from a single qubit to a register of $n$ separable, non-interacting qubits. Each local qubit state is represented by a $2\times2$ density matrix,
\begin{equation}
    \rho^{(t,k)}_i =
    \begin{pmatrix}
        \alpha_{1,i}^{(k,t)} & \beta_{1,i}^{(k,t)} + i\beta_{2,i}^{(k,t)} \\
        \beta_{1,i}^{(k,t)} - i\beta_{2,i}^{(k,t)} & \delta_{1,i}^{(k,t)}
    \end{pmatrix},
\end{equation}
where Hermiticity implies that only four real parameters are required in practice. The full separable register is then written as
\begin{equation}
    \tilde{\rho}^{(t)}_i = \bigotimes_{k=1}^{n} \rho^{(t,k)}_i .
\end{equation}
Likewise, the local noisy evolution is also separable,
\begin{equation}
    \mathcal{M}(\rho_i^{(t)})
    =
    \mathcal{M}_1(\rho_i^{(t,1)})\otimes
    \mathcal{M}_2(\rho_i^{(t,2)})\otimes \cdots \otimes
    \mathcal{M}_n(\rho_i^{(t,n)}),
\end{equation}
Because the local weak measurements are sampled independently, in general, $\mathcal{M}_i \neq \mathcal{M}_j$. As each local Hermitian density matrix is specified by four real numbers, the dataset has the dimensions:
\begin{equation}
    [B,S,n,4]\;\rightarrow\;[B,S,4n].
\end{equation}
Thus, the number of features grows linearly with the number of qubits, in contrast to the exponential scaling associated with a general entangled density matrix. This makes the separable-register task suitable for sequence-based architectures such as the LSTM.

In this setting, the LSTM model successfully learns the inverse local trajectories. An example of a six-qubit separable quantum register is shown in Fig.~\ref{fig:6 qubits}. The model receives the final noisy output of the forward process and reconstructs the corresponding pre-noise state with $\mathcal{F}_{GM}>0.99$. This result demonstrates that, for separable registers subject to local noisy channels, the inverse reconstruction problem remains tractable and retains linear scaling with the number of qubits.

The observed high reconstruction $\mathcal{F}_{GM}$ indicates that LSTM-based models are sufficient for calibrating independent or weakly correlated qubits undergoing local noise. However, this favorable behavior relies on the separability of the register. Once non-local correlations are introduced, the relevant state representation grows exponentially with the number of qubits, and more expressive architectures are required.

\subsection{Entangled multi-qubit datasets}

\begin{figure*}[t]
    \centering

    \begin{subfigure}[t]{0.33\textwidth}
        \centering
        \includegraphics[width=\linewidth]{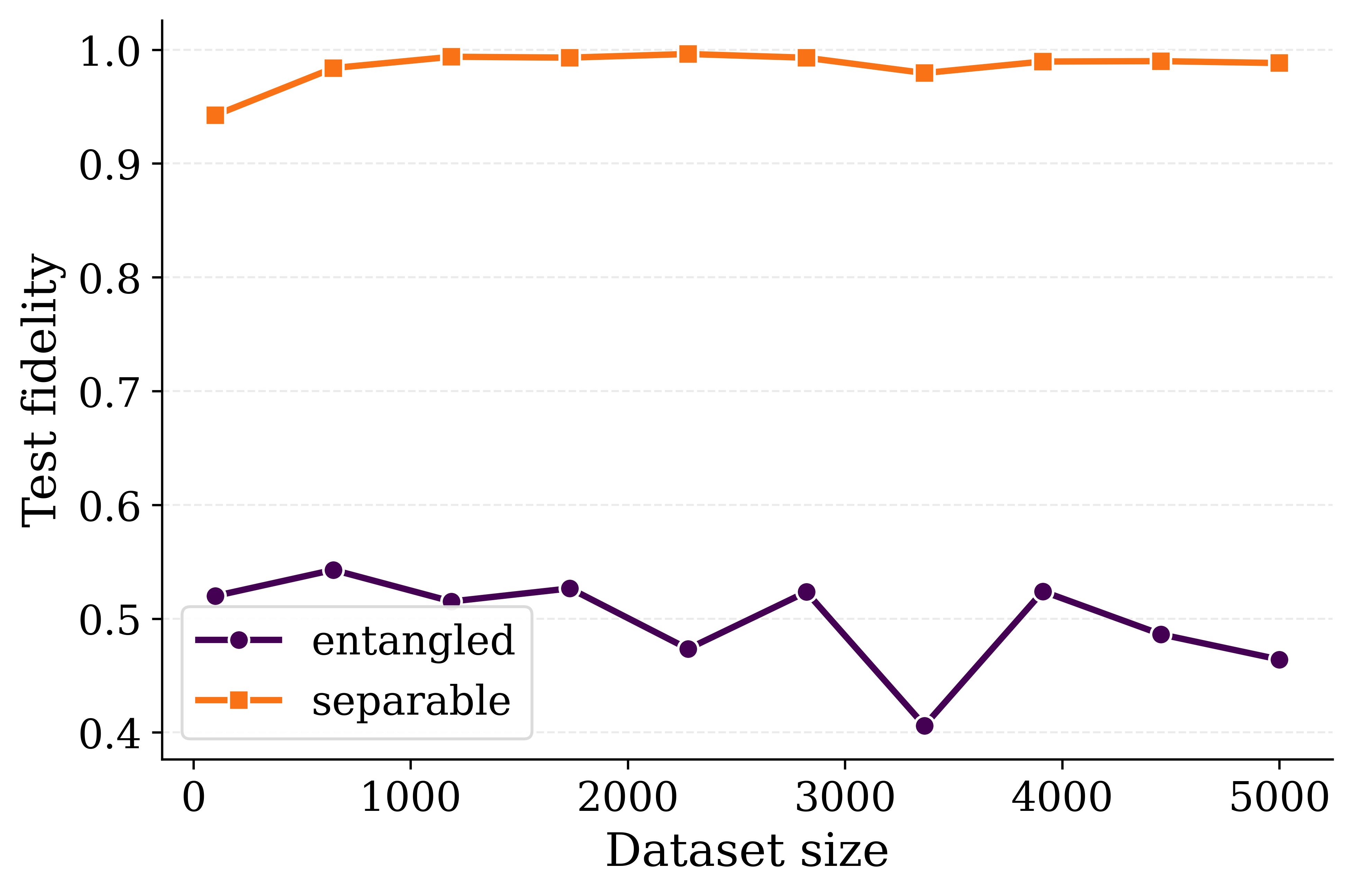}
        \caption{}
        \label{fig:lstm-datasize}
    \end{subfigure}
    \hfill
    \begin{subfigure}[t]{0.33\textwidth}
        \centering
        \includegraphics[width=\linewidth]{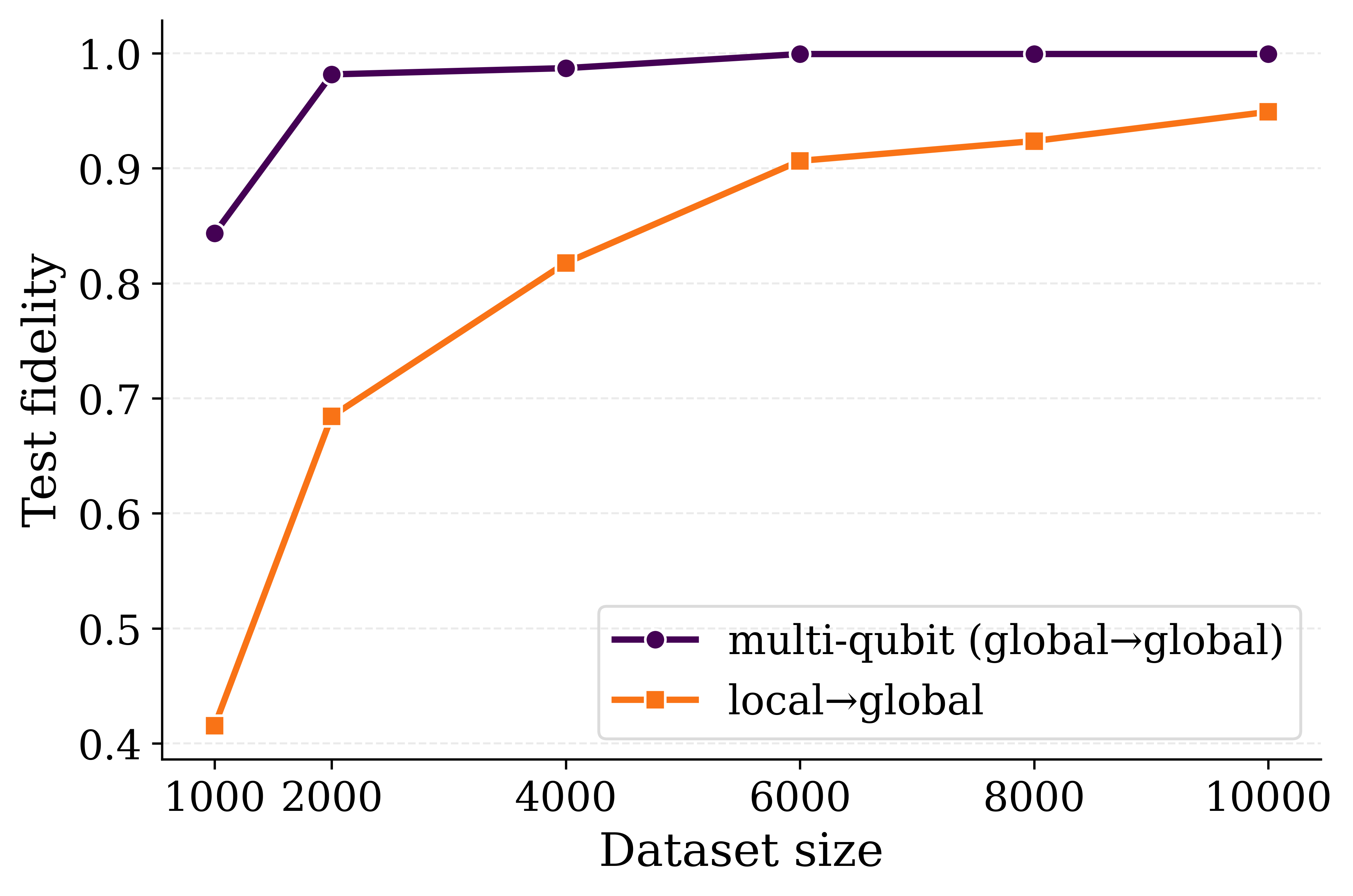}
        \caption{}
        \label{fig:diffusion-datasize}
    \end{subfigure}
    \hfill
    \begin{subfigure}[t]{0.32\textwidth}
        \centering
        \includegraphics[width=\linewidth]{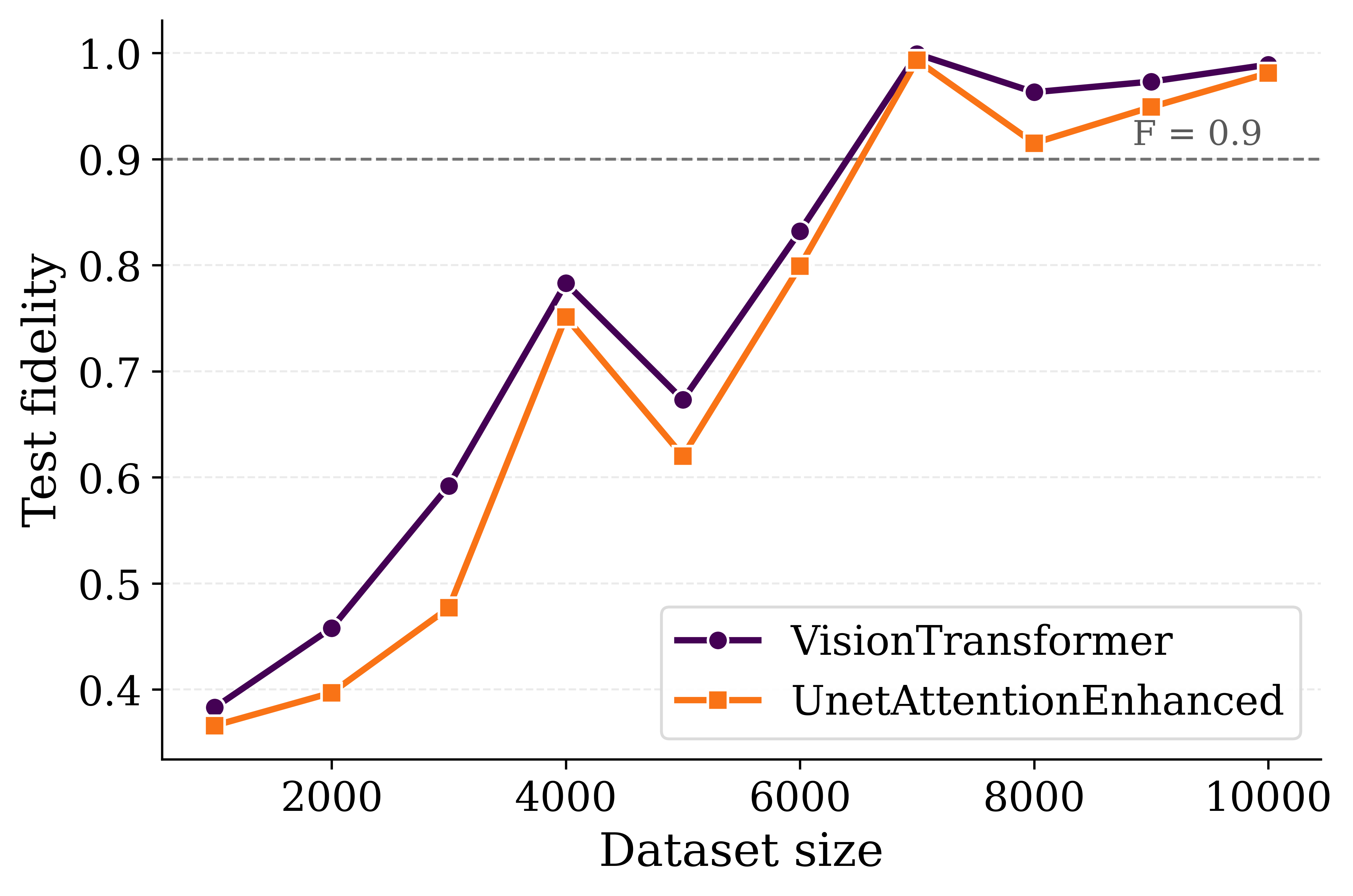}
        \caption{}
        \label{fig:benchmarks}
    \end{subfigure}

\caption{Test fidelity as a function of training-set size for different four-qubit reconstruction tasks and model architectures. (a)~The same LSTM architecture trained on entangled states (full $d\times d$ density matrix, purple circles) versus separable product states (a collection of $n$ single-qubit $2\times2$ matrices, orange squares), with $\epsilon=0.1$ and $\eta=0.55$. The separable representation is learned almost perfectly from as few as $100$ examples, whereas the fidelity for entangled states remains near $0.5$ regardless of dataset size, highlighting the difficulty of learning genuinely entangled states. (b)~The same ViT diffusion model trained to denoise the full $16\times16$ global density matrix directly (purple circles) versus reconstructing the global state from noisy single-qubit marginals (orange squares), with $\epsilon=0.1$ and $\eta=0.6$. Each point is the mean over 10 independent runs. The global-to-global model saturates near unit fidelity with only ${\sim}2{,}000$ examples, whereas the local-to-global model reaches $0.949$ at $10^4$ examples. (c)~Comparison between the Vision Transformer and enhanced U-Net for the four-qubit local-to-global reconstruction task, with $\epsilon=0.15$ and a purity threshold of $0.55$. The Vision Transformer consistently outperforms the enhanced U-Net across the tested dataset sizes, while both architectures exceed $\mathcal{F}_{GM}=0.95$ for sufficiently large training sets.}
    \label{fig:fidelity-vs-datasize-both}
\end{figure*}

We now turn to entangled quantum registers. In this setting, the global quantum-register state belongs to a Hilbert space of dimension $2^n$. The information of this dataset is represented by $d\times d$ complex density matrices, corresponding to $2d^2=2\cdot4^n$ stored real values. This exponential scaling makes the learning task substantially more demanding than in the separable cases discussed above. As a result, LSTM models exhibit poor performance on such systems, as shown in Fig.~\ref{fig:lstm-datasize}. Consequently, the computational time and resources required to learn the non-local structure of the entangled states became impractical as the system size increased.

We therefore deploy attention-based models, namely the Vision Transformer and the enhanced U-Net, to learn the full non-local structure of the density matrix. To generate the entangled dataset, we first prepare random product states and then apply $m=n-1$ randomly chosen CNOT gates between qubits, thereby creating a non-separable quantum register. The same sequential weak-measurement channel is then applied to the entangled state. The corresponding dataset has the structure
\begin{equation}
[B,S,2,d,d]\rightarrow[B,S,2,d,d],\qquad d=2^n.
\end{equation}

We consider two learning settings in this regime. In the first \textit{global-to-global} setting, the model receives the full noisy density matrix and predicts the corresponding full denoised density matrix. This task tests whether the model can learn an effective inverse map when the complete non-local state representation is available. In the second \textit{local-to-global} setting, the model only receives as its input the local reduced density matrices
\begin{equation}
\rho_i = \mathrm{Tr}_{j\neq i}(\rho),
\end{equation}
while the target remains the full global density matrix. The corresponding compressed input has shape
\begin{equation}
[B,S,n,2,2,2],
\end{equation}
and therefore omits explicit information about inter-qubit correlations.

Both settings are overviewed briefly in Fig. \ref{fig:paper's_pipline}, with a more detailed overview of the local-to-global setting in Fig.~\ref{fig:local_to_global}. The local-to-global approach is much more experimentally feasible, as it relies only on local measurements whose number scales linearly with the register size, rather than on the exponentially many measurements required for full quantum-state tomography \cite{aaronson2018shadow, huang2020predicting}. However, this local-to-global reconstruction task is also significantly more challenging, as the model must infer missing many-body correlations from local marginals alone. Representative results for the local-to-global reconstruction task are shown in Figs.~\ref{fig:sample_loc}  and~\ref{fig:test_fidelity_hist}, respectively, while Fig~\ref{fig:diffusion-datasize} illustrates the relative difficulty of the local-to-global task vs the global-to-global task in terms of model performance and convergence time.

The results demonstrate a clear separation among the learning regimes considered in this work, highlighting the need for more advanced models capable of capturing complex correlations while maintaining computational efficiency. In particular, these architectures handled the seven-qubit setting more effectively than the LSTM. We expect that improved data management and further optimization could enable the framework to scale to even larger quantum systems. Likewise, for larger systems, standard regularization methods, data augmentation, and architectures designed to exploit physical structure may be necessary to improve generalization.

Figure~\ref{fig:benchmarks} further compares the performance of the ViT and enhanced U-Net architectures on four-qubit local-to-global datasets with five-step sequences, presenting the output $\mathcal{F}_{GM}$ as a function of dataset size. The ViT consistently outperforms the enhanced U-Net across all tested dataset sizes. This advantage may be attributed to its self-attention mechanism, which enables efficient parallel processing and facilitates the identification of complex, long-range correlations within the density-matrix representation. Nevertheless, the comparison also highlights an important scalability challenge: as the number of qubits increases, the dimension of the target state, the complexity of the noise process, the required sequence length, and the amount of requisite training data all grow rapidly. These trends stem from the fundamental complexity of quantum systems and emphasize that classical machine-learning models can support quantum computation by assisting in calibration, reconstruction, and error mitigation, but do not remove the fundamental scaling challenges associated with quantum many-body systems. Rather than providing global correction to a monolithic quantum device, such models should be deployed in a modular manner on tiled and distributed quantum-computing architectures, where quantum information is transmitted through noisy, context-dependent channels.

\section{Discussion}\label{sec:dis}
In this manuscript, we have introduced a quantum analogue of a classical diffusion process, in which sequential weak measurements generate a diffusion-like noisy trajectory for quantum states. This construction adds more flexibility and nuance than merely using an ideal global depolarizing channel, as individual weak-measurement trajectories consist of basis-dependent local dephasing steps, while the ensemble-averaged dynamics are locally depolarizing. In this framework, local weak measurements induce a non-unitary forward evolution that gradually degrades the input state. Machine-learning models are then trained on the resulting trajectories to approximate an effective inverse map, thereby reconstructing the corresponding pre-noise state and prescribing error mitigation.

The proposed methodology was tested on several datasets of increasing complexity. For single-qubit and separable multi-qubit registers, where the number of features scales linearly with the number of qubits, LSTM-based models were sufficient to learn the inverse trajectories. For entangled quantum registers, where the density-matrix representation scales exponentially, attention-based architectures such as the Vision Transformer (ViT) and enhanced U-Net were required. We also considered a local-to-global reconstruction task, in which the model receives only local reduced density matrices while the target is the full global density matrix. This setting was more challenging and more prone to overfitting, since the model must infer non-local correlations that are not explicitly present in the input.

The main practical limitation of the present implementation is the rapid growth of classical computational resources with the number of qubits. Larger systems require larger datasets, longer training times, and substantially more memory, especially when full entangled density matrices are used. Nevertheless, the results show that trajectory-based learning can provide a useful tool for AI-assisted quantum-error mitigation. In particular, the separable-register setting suggests that such models may be useful for calibrating local noisy channels before computation, while the entangled and local-to-global settings demonstrate the potential of attention-based models for more complex quantum-state reconstruction tasks, whether local or modular.

Future work may extend the proposed diffusion framework to state-preparation protocols or embed the learned inverse process into circuit-based architectures with redundant qubits. Since the sequential weak-measurement dynamics define a contextual, non-unitary process, the present method may also be viewed as a restricted form of process-level learning. This suggests a possible route toward learning, characterizing, and mitigating broader classes of noisy quantum processes. Such tools may be especially relevant for interconnected and distributed quantum-computing architectures, where quantum information is transmitted through noisy and context-dependent channels. When leveraged for such applications, the techniques developed in this work expand AI models' role as complementary tools for improving and scaling quantum hardware.

\section*{Acknowledgements}
This work was supported by the European Union's Horizon Europe research and innovation programme under grant agreement No. 101178170, by the Israel Science Foundation under grant agreement No. 2208/24 and by the Israeli Council for Higher Education through ``QERNEL'' and the ``Interdisciplinary Center for the Theory of Quantum Computing''.

\section*{Author contributions}
Y.I., E.M., T.L.P. and E.C. conceived and initiated the study. All authors
developed the methodology and designed the research. Y.I. and E.C.
developed the theoretical framework. Y.I., O.N. and T.L.P. developed the
algorithm, implemented, trained, and evaluated the diffusion and
transformer models. Y.I. and E.M. secured and allocated the computational
resources. Y.I. and O.N. performed the computational investigations,
validated the models, analyzed the results, and prepared the figures.
Y.I. wrote the original draft. All authors reviewed and edited
the manuscript. E.M., T.L.P. and E.C. supervised the research. All authors
discussed the results and approved the final manuscript.

\section*{Competing interests}
The authors declare no competing interests.

\section*{Additional Information}
\subsection*{Data availability}
The data supporting the findings of this study are provided within the article.  Additional data and materials are available from the corresponding author upon reasonable request.
\subsection*{Code availability}
The custom code and trained model parameters developed in this study are not publicly available at present because they are subject to ongoing intellectual-property protection and commercialization procedures. Requests for access may be directed to the corresponding author and will be considered on a case-by-case basis.

\bibliographystyle{ieeetr}
\bibliography{references.bib}

\appendix

\section{Neural-network architectures and training methodology for error mitigation}\label{app:models}

To learn the inverse noisy evolution, we consider three neural-network architectures adapted to different data regimes. For relatively small sequential datasets, we employ a Long Short-Term Memory (LSTM) network, which naturally captures temporal correlations along the inverse trajectory. For larger multi-qubit datasets, whose representation grows rapidly with system size, we use two image-based architectures: a Vision Transformer (ViT) \cite{touvron2021training,dosovitskiy2020image,wu2021cvt} and a custom enhanced U-Net \cite{ronneberger2015u,isensee2021nnu,oktay2018attention}. The ViT is used as a transformer-based architecture with a fixed patch-token structure, while the enhanced U-Net combines multiscale feature extraction with global context modeling. In all cases, the goal is to approximate the inverse evolution and reconstruct the initial state from the final noisy state. After the attention-based architectures, we apply a density-matrix filter to ensure that the model output is a valid density matrix; see Appendix~\ref{Appenix:density_filter}.

\subsection{LSTM}
To approximate the inverse quantum channel in Eq.~\ref{eq:inverse}
we train an LSTM network to map noisy quantum trajectories back to their original pure states. In an LSTM, each cell state $c_n$ stores information accumulated up to time step $n$ and is updated together with the hidden state $h_n$ to produce the next pair $(c_{n+1},h_{n+1})$. Propagating these internal states through the sequence enables the model to capture temporal correlations and long-range dependencies.

The training data are constructed by reversing the simulated forward trajectories, such that the target at each step corresponds to the previous, less noisy state rather than the next, noisier one. The task is therefore formulated as a sequential regression problem in which the model predicts the inverse evolution step by step in an autoregressive manner \cite{uria2016neuralautoregressivedistributionestimation,graves2014generatingsequencesrecurrentneural}; see Appendix~\ref{Appendix:teacher}. The implemented architecture consists of a multi-layer LSTM encoder followed by a residual feedforward head that maps the latent representation to the predicted density matrix.

To stabilize training, we employ a teacher-forcing scheduler \cite{sutskever2014sequence}, whereby the model receives either the ground-truth input or its own previous prediction according to a probability that decays during training. After training, the model receives the final noisy state $\rho^{(T)}$ and predicts an approximate inverse trajectory,
\begin{equation}
    \rho^{(T)} \mapsto \rho^{(T-1)} \mapsto \cdots \mapsto \rho^{(0)}.
\end{equation}

A limitation of LSTMs is their inherently sequential processing across time steps, which restricts parallelization and may slow both training and inference on modern hardware. This motivates our use of attention-based image architectures for larger quantum datasets.

\subsection{Vision Transformer}
The ViT \cite{touvron2021training,dosovitskiy2020image,wu2021cvt} is more suitable for larger density-matrix representations,  which treats the input as a multi-channel image and processes it using self-attention. This representation is particularly useful when the relevant quantum correlations are spatially non-local in the density-matrix layout.

The complex-valued input is first encoded as a multi-channel image and divided into non-overlapping square patches. These patches are linearly projected into token embeddings, to which fixed two-dimensional sinusoidal positional encodings are added. The resulting token sequence is then processed by a stack of Transformer encoder blocks, each composed of multi-head self-attention, a feedforward MLP, residual connections, and pre-normalization LayerNorm \cite{vaswani2023attentionneed}.

To reconstruct the output, the processed tokens are reshaped back onto the spatial lattice and passed through a lightweight projection head followed by a final $1\times1$ convolution, yielding the denoised multi-channel output. During training, the Vision Transformer is optimized to directly regress from the noisy representation to the target denoised state. In contrast to the LSTM, which explicitly models the inverse process as a sequence, the Vision Transformer learns a global mapping between the corrupted input and the reconstructed output.

Compared with recurrent architectures, self-attention enables greater parallelism and captures long-range dependencies uniformly across the full field. Empirically, this improves scalability for larger systems and provides stable performance. An illustration of the implemented Vision Transformer is shown in Fig.~\ref{fig:vit}.

\subsection{Enhanced U-Net}

As a second image-based model, we employ an enhanced U-Net architecture \cite{ronneberger2015u,isensee2021nnu,oktay2018attention}. Whereas the Vision Transformer emphasizes global self-attention, the U-Net is designed to combine local feature recovery with multiscale contextual reconstruction. This is advantageous for denoising tasks in which both fine local structure and global organization of the density matrix are important.

Our enhanced U-Net follows an encoder--attention--decoder architecture. The encoder maps the noisy quantum trajectory into a latent feature representation by progressively reducing the spatial resolution while increasing the number of feature channels through residual blocks. Downsampling is performed using average pooling rather than max pooling, since average pooling preserves the mean activation within each pooling window and therefore retains more of the global signal structure. An attention module is then applied at the bottleneck to identify the most informative feature correlations for the inverse reconstruction task.

\begin{figure}[t]
    \centering
    \begin{subfigure}[b]{1\linewidth}
        \centering
        \includegraphics[width=\linewidth]{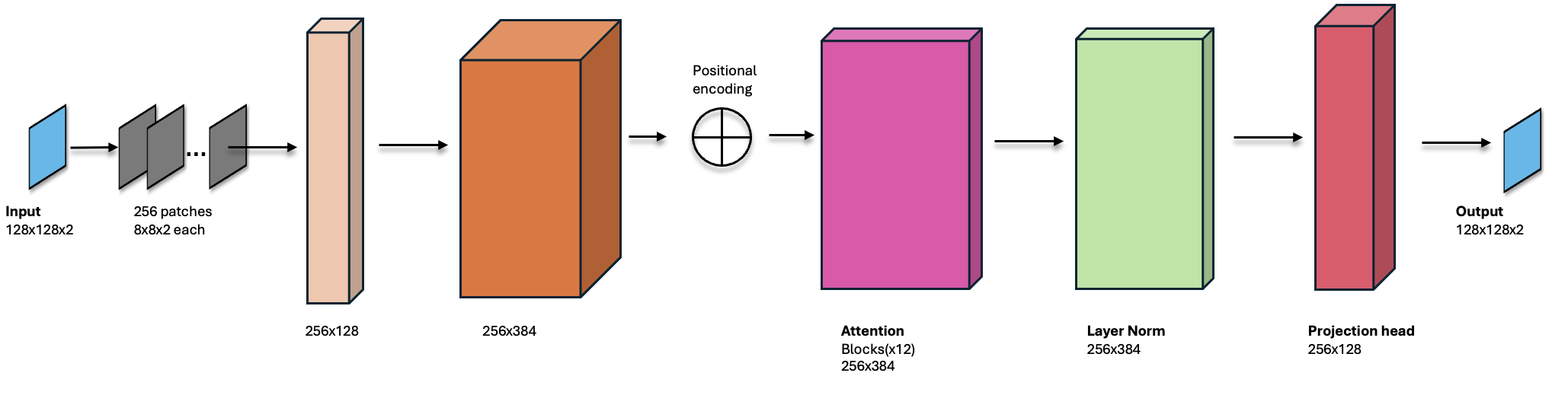}
        \caption{Vision Transformer architecture.}
        \label{fig:vit}
    \end{subfigure}
    \hfill
    \begin{subfigure}[b]{1\linewidth}
        \centering
        \includegraphics[width=\linewidth]{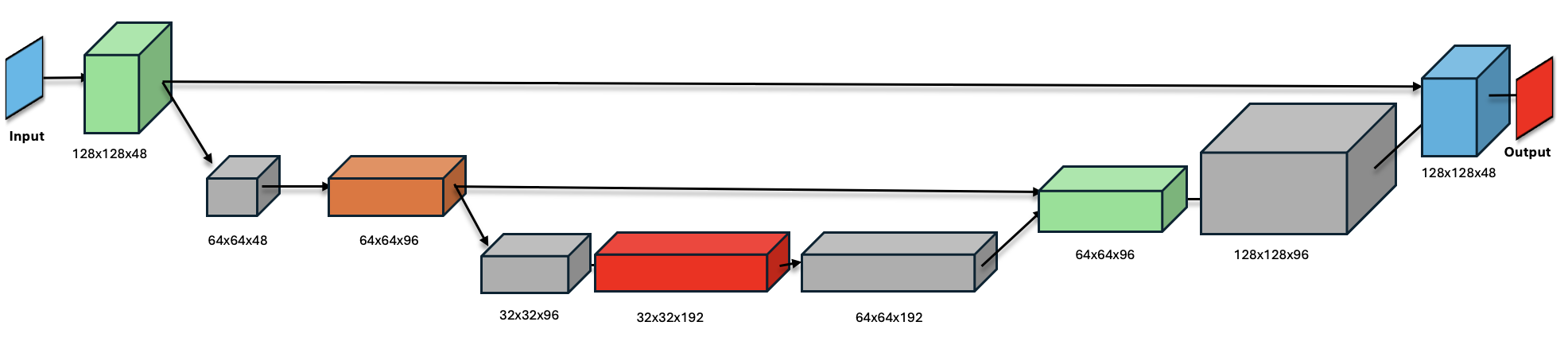}
        \caption{Enhanced U-Net architecture.}
        \label{fig:enhencetunet}
    \end{subfigure}
    \caption{Schematic architectures of (a) the Vision Transformer and (b) the enhanced U-Net used in this work.}
    \label{fig:vit_unet_subfig}
\end{figure}

The decoder mirrors the encoder by progressively upsampling the latent representation back to the original resolution. Skip connections between corresponding encoder and decoder stages help preserve fine-scale structural information, while the bottleneck attention layer provides global contextual features. During training, the enhanced U-Net is optimized as a direct denoiser, mapping the noisy density-matrix representation to its reconstructed counterpart. The full architecture is illustrated in Fig.~\ref{fig:enhencetunet}.

\section{Teacher Forcing and autoregression}\label{Appendix:teacher}
We adopt a training strategy that begins with teacher forcing and transitions gradually to autoregressive generation. In the early stages of training, the model receives the ground-truth target from the previous time step as input at each point in the sequence. This accelerates convergence by exposing the model to ideal  local inputs. However, reliance on ground truth during training leads to a mismatch at inference time, when only the model’s own predictions are available.
To bridge this gap, we use a linear-plateau schedule to anneal the teacher forcing probability. Over the course of training, the model increasingly relies on its own outputs from previous steps, adapting to the eventual test-time scenario. This balance between teacher forcing and autonomous prediction improves robustness and stability.

\section{Density matrix filter}\label{Appenix:density_filter}

A post-processing filter is applied to enforce the density matrix structure. 
Given a reconstructed matrix $\rho$, the closest physical state is obtained by projecting onto the set of valid density operators, i.e., positive semidefinite matrices with unit trace. 
This projection is performed via spectral decomposition,
\begin{equation}
    \rho = V \Lambda V^\dagger,
\end{equation}
followed by eigenvalue thresholding $\lambda_i \mapsto \max(\lambda_i,0)$ to enforce positivity, and normalization to ensure $\mathrm{Tr}(\rho)=1$. 
This procedure corresponds to the Euclidean (Frobenius norm) projection onto the convex set of density matrices, and is widely used in quantum state tomography \cite{Smolin2012}, with its mathematical foundation in nearest positive semidefinite matrix approximations \cite{Higham1988} and convex optimization theory \cite{Boyd2004}. 
The computational cost is dominated by the eigendecomposition, scaling as $\mathcal{O}(N^3)$ for an $N \times N$ matrix, making the procedure efficient for moderate system sizes.

\end{document}